\documentclass[useAMS,usenatbib]{mn2e}
\usepackage{psfig}

\newcommand{\Msun}{ M_{\odot}}
\newcommand{\Mdot}{\dot{M}}
\newcommand{\mdot}{\dot{M_1}}
\newcommand{\Mbh}{M}
\newcommand{\teff}{T_{\rm eff, max}}

\title[Cold Accretion Disks and Lineless Quasars]{Cold Accretion Disks and Lineless Quasars}
\author[Ari Laor and Shane W. Davis]{Ari Laor$^{1}$\thanks{E-mail:
laor@physics.technion.ac.il (AL); swd@cita.utoronto.ca (SWD)} and Shane W. Davis$^{2}$\footnotemark[1]\\
$^{1}$Physics Department, Technion, Haifa~32000, Israel\\
$^{2}$Canadian Institute for Theoretical Astrophysics. Toronto, ON M5S3H4, Canada}
\begin{document}

\date{}

\pagerange{\pageref{firstpage}--\pageref{lastpage}} \pubyear{2002}

\maketitle

\label{firstpage}

\begin{abstract}

The optical-UV continuum of quasars is broadly consistent with the emission from a 
geometrically thin
optically thick accretion disk (AD). The AD produces the ionizing 
continuum which powers the broad and narrow emission lines. 
The maximum AD effective temperature is given by 
\mbox{$\teff=f_{\rm max}(\Mdot/M^2)^{1/4}$}, where $M$ is the black hole mass,
$\Mdot$ the accretion rate, and $f_{\rm max}$ is set by the black hole spin $a_*$.
For a low enough value of $\Mdot/M^2$ the AD may become too cold to produce ionizing photons.
Such an object will form a lineless quasar. This occurs for a local blackbody (BB) AD with a 
luminosity $L_{\rm opt}=10^{46}$~erg~s$^{-1}$
for $M>3.6\times 10^9\Msun$, when $a_*=0$, 
and for $M>1.4\times 10^{10}\Msun$, when $a_*=0.998$.
Using the AD based $\Mdot$, derived from $M$ and $L_{\rm opt}$, and the
reverberation based $M$, derived from $L_{\rm opt}$ 
and the H$\beta$ FWHM, $v$, gives $\teff\propto L_{\rm opt}^{-0.13}v^{-1.45}$. 
Thus, $\teff$
is mostly set by $v$. Quasars with a local BB AD become lineless for
$v> 8,000$~km~s$^{-1}$, when $a_*=0$, and for $v> 16,000$~km~s$^{-1}$,
when $a_*=0.998$. Higher values
of $v$ are required if the AD is hotter than a local BB.
The AD becoming non-ionizing may explain why line emitting quasars with $v>10,000$~km~s$^{-1}$ 
are rare. 
Weak low ionization lines may still be present if the X-ray continuum is luminous
enough, and such objects may form a population of weak emission line
quasars (WLQ). If correct, such WLQ should show a steeply falling SED at $\lambda<1000$\AA. 
Such an SED was observed by Hryniewicz et al. in SDSS J094533.99+100950.1, 
a WLQ observed down to 570\AA, which is well modeled by a rather cold AD SED.
UV spectroscopy of $z\sim 1-2$ quasars is required to eliminate
potential intervening Lyman limit absorption by the intergalactic medium (IGM), 
and to explore if the SEDs of lineless quasars and
some additional WLQ are also well fit by a cold AD SED.

\end{abstract}


\section{Introduction}

The optical-UV spectral energy distribution (SED) of quasars is broadly consistent
with the expected emission of a thin accretion disk (AD) around a massive black hole
(Shields 1978; Czerny \& Elvis 1987; Sun \& Malkan 1989; Laor 1990; Blaes et al. 2001; 
Shang et al. 2005).
Various spectral features, such as the general turnover at 
$\lambda<1000$\AA\ (Zheng et al. 1997; Telfer 2002; Barger \& Cowie 2010),
the observed optical-UV spectral slopes (Bonning et al. 2007; Davis et al. 2007), 
the small dispersion in the UV/optical flux ratio (Davis \& Laor 2011),
and microlensing variability 
(Morgan et al. 2010; Blackburne et al. 2011), may imply various 
modifications beyond the simplest AD models 
(see review by Koratkar \& Blaes 1999, and citations thereafter).
One should note that some of the expected AD spectral features may be diluted by 
various unrelated emission components (Kishimoto et al. 2004; 2008).

The peak of the simplest AD model of local blackbody (BB) emission occurs at
$\nu_{\rm peak} \propto (\Mdot/M^2)^{1/4}$ (see \S 2), 
or equivalently $\nu_{\rm peak}\propto (l/M)^{1/4}$, where
$M$ is the black hole mass, $\Mdot$ is the accretion rate, and
$l=L/L_{\rm Edd}$ is the ratio of the bolometric luminosity to the Eddington luminosity. 
Thus, when $M$ increases from $10\Msun$ to $10^9\Msun$,
at a fixed $l$, the disk cools by a factor of 100. This factor is consistent with the
drop from $\nu_{\rm peak}\simeq 1-2$~keV in X-ray binary systems
(e.g. Remillard \& McClintock 2006), 
to $\nu_{\rm peak}\simeq 10-20$~eV, observed in quasars. The quasar SED 
peaks at about the H ionization threshold, and thus provides ample ionizing luminosity
which powers the broad and narrow line emission. 

A further extrapolation of the expected AD SED to $M>10^9\Msun$ and $l<0.1$ leads to 
$\nu_{\rm peak}<10$~eV and thus to a non-ionizing AD SED
(e.g. see Laor \& Netzer 1989 fig.5, for a model where
$\nu_{\rm peak}\simeq 1-2$~eV). Such models appeared inconsistent with the
observations, as the SED of such quasars cannot power strong line emission. 
Such quasars will appear mostly as luminous continuum sources. Some line emission
can still be produced through X-ray photoionization. However, in luminous quasars the
fraction of the bolometric luminosity in the x-ray drops well 
below 10\% (Just et al. 2007), and the fraction of the bolometric power 
available for line emission thus drops from $>50$\% to $<10$\%., i.e.
by a factor of $>5$. The total line emission should drop by a similar factor.

Observational evidence that weak line quasars (WLQ) do exist started to accumulate since the
discovery of McDowell et al. (1995) that in PG~1407+265 all lines, except H$\alpha$, 
are exceptionally weak. A BL Lac origin could be clearly excluded 
based on its SED which overlaps well the mean SED of optically selected 
quasars (Elvis et al. 1994). Additional WLQ were discovered in various other 
studies (Fan et al. 1999, 2006; Anderson et al. 2001; Hall et al. 2002, 2004; 
Reimers et al. 2005; Ganguly et al. 2007; Leighly et al. 2007a; Hryniewicz et al. 2010), 
and were followed up by dedicated studies of larger samples of WLQ at a range of wavelengths 
(Shemmer et al. 2009, 2010; Diamond-Stanic et al. 2009; Plotkin et al. 2010a, 2010b; 
Wu et al. 2011).  
The observed radio to X-ray SEDs of most WLQ is consistent with the mean SED of
radio quiet quasars, and clearly excludes the BL Lac interpretation for the absence of 
line emission. Proposed explanations for WLQ
invoked unusual Broad Line Legion (BLR) properties, such as a low covering factor, an anisotropic
ionizing source, or BLR shielded from the ionizing source.

A possibly related issue is that of `true type 2' AGN, i.e. AGN which lack
broad lines, but appear to be unobscured (Tran 2003; Shi et al. 2010).
These objects differ from WLQ in two ways. First, they are defined by the lack of broad
lines, and not by the lack of lines in general. However, once $z\ga 1$, the significant
narrow lines are redshifted from optical spectroscopy, and one cannot separate
a lineless quasar from a type 2 AGN. Second, the `true type 2' AGN 
tend to reside at $l<10^{-2}$ (Shi et al. 2010; Trump et al. 2011), in contrast 
with WLQ which typically have $l\sim 0.1-1$ (e.g. Shemmer et al. 2010).

Various suggestions were made to explain the lack of broad lines in unobscured type 2 AGN.
Nicastro (2000), and Nicastro et al. (2003), suggested that
the BLR is formed by accretion disk instabilities occurring in a critical radius 
where the disk changes from gas pressure dominated to radiation pressure dominated. This 
critical radius can become smaller than the innermost stable orbit, and 
then the BLR cannot form. Czerny et al. (2004) and Elitzur \& Ho (2009) 
suggested that the transition
of the AD from a cold geometrically thin flow, to a hot advection dominated accretion flow, 
eliminates the AD wind which launches the BLR.
Elitzur \& Shlosman (2006) suggested that the BLR is formed by a torus outflow, and
below a certain accretion rate the outflow transforms into radio jets, and the BLR 
disappears. The various physical mechanisms suggested above to eliminate the BLR 
generally occur below some values of $l$, which may be a function of $M$. These ranges
correspond to a certain maximum line width, which may be function of $L$, beyond which
all AGN become `true type 2' AGN. Laor (2003) pointed out that the BLR appears not to be
detectable at FWHM $> 25,000$~km~s$^{-1}$, which may either be a
detection limit, or a maximal velocity where the BLR can exist. However, no physical mechanism
was put forward to explain why the BLR FWHM may be the primary parameter for the existence
of the BLR. As shown below, 
a non-ionizing AD may provide a simple explanation for the observed luminosity independent 
upper limit on the BLR FWHM.

In the following section we derive the accretion disk model parameters 
which produce a cold AD, i.e. an AD with a non-ionizing SED. In \S 3 we
show that the SED of a WLQ, observed as far as 570\AA, is fit
surprisingly well by a cold AD model. In \S 4 we discuss additional implications, and the
relevance to other observed properties of AGN. The main conclusions are summarized in \S 5.

\section{Non-ionizing AD model parameters}

Below we first derive the relation between $\teff$ and the accretion disk model
parameters, based on the standard Shakura \& Sunyaev (1973, SS73) thin disk model.
The flux emitted per unit area is
\begin{equation}
 F=\frac{3}{8\pi}\frac{G\Mdot \Mbh}{R^3}f(R,M,a_*)
\end{equation}
where $R$ is the radius, and $f(R,M,a_*)$ is a dimensionless factor
set by the inner boundary condition, 
and the relativistic effects (Novikov \& Thorne 1973; Riffert \& Herold 1995).
It is convenient to use the dimensionless radius, $r\equiv R/R_g$,
where $R_g\equiv G\Mbh/c^2$, which gives
\begin{equation}
 F=\frac{3c^6}{8\pi G^2}\frac{\Mdot}{\Mbh^2 r^3}f(r,a_*) ,
\end{equation}
where $f(r,a_*)\la 1$.

The local effective temperature is $T_{\rm eff}\equiv (F/\sigma)^{1/4}$,
and it scales as  
\begin{equation}
T_{\rm eff} = T_0 f(r,a_*)^{1/4}r^{-3/4} , 
\end{equation}
where
\begin{equation}
T_0 \equiv \left(\frac{3 c^6}{8\pi G^2 \sigma} \right)^{1/4}
\frac{\Mdot^{1/4}}{\Mbh^{1/2}}. ,
\end{equation}  
or in more convenient units
\begin{equation}
T_0=8.6\times 10^5\mdot^{1/4}m_8^{-1/2}~{\rm K}
\end{equation} 
where 
\[
\Mdot=\mdot~ \Msun~yr^{-1},
\] 
and
\[
 M=10^8m_8~\Msun ,
\]
or in cgs units 
\[ 
\Mdot=6.3\times 10^{25}\mdot~{\rm gr~s}^{-1},
\]
and  
\[
M=1.989\times 10^{41}m_8~{\rm gr} .
\]
We denote by $f_{\rm max}$ the maximum value of $f(r,a_*)$,
which sets the maximum disk temperature
$\teff$, assuming a local BB emission. The ratio 
$\teff/T_0$ is then a function of $a_*$ only. Ratios
for representative values of $a_*$ are provided in Table 1.
As $a_*$ increases, the innermost marginally stable disk radius $r_{\rm ms}$ decreases,
and $\teff/T_0$ increases.

Note that $f_{\rm max}$ is only a function of $a_*$ as we have assumed
the standard AD structure, where the viscous torque increases inwards,
reaches a maximum, and then drops until it vanishes at the innermost
stable circular orbit (ISCO).  In reality, there is likely to be some
magnetic stress at or near the inner boundary and continuing emission
inside the ISCO.  Quantifying the level of additional emission from
near or inside the ISCO is an active research topic (e.g Kulkarni
  et al. 2011, Noble et al. 2011).  However, any additional emission
is not likely to cause significantly greater variation in $f_{\rm
  max}$ than varying $a_*$ alone.  For example, Noble et
al. (2011) generally argue for a larger effect due to emission inside
the ISCO than Kulkarni et al. (2011).  But they find that an $a_*=0$
would appear as an $a_*=0.2-0.3$ model if the emission at the disk
surface is locally everywhere a BB. (Note, however, that the radiation
inside the ISCO may be far from thermodynamic equilibrium due to the
drop in disk surface density, so local BB emission may not be a good
approximation there.)  Hence, in this work we will focus on the simple AD
  models where $f_{\rm max}$ is determined solely by $a_*$.

What is the frequency $\nu_{\rm max}$ where the accretion disk $\nu L_{\nu}$
peaks?  For a single temperature BB, it is simple to show that
the peak occurs at $h\nu_{\rm max}/kT_{\rm max}=3.92$. 
A local BB AD is a superposition of blackbodies with $T\leq T_{\rm max}$, 
weighted by the emitting 
surface area, and convolved by the relativistic effects. This superposition 
results in an integrated spectrum with a broader spectral energy distribution (SED)
compared to a single temperature BB, which peaks at 
$h\nu_{\rm max}/kT_{\rm max}<3.92$.
Table 1 lists the values of $h\nu_{\rm max}/kT_{\rm max}$, 
corresponding to an AD observed at 
$\mu\equiv \cos\theta=0.8$, where $\theta$ is the inclination angle of the AD to the 
line of sight ($\mu=1$ is face on).

How low should $\nu_{\rm max}$ be for an AD to be called non-ionizing?
Let us define a non-ionizing AD, or a cold AD, as an AD where the ionizing luminosity is 
$\le 0.01$ of the bolometric luminosity.
We now look for $\nu_{0.01}$, the frequency above which the
integrated luminosity is 1\% of the bolometric luminosity, i.e.
\[ 
\int_{\nu_{0.01}}^\infty L_{\nu}d\nu/\int_0^\infty L_{\nu}d\nu=0.01 . 
\]
If $\nu_{0.01}<3.29\times 10^{15}$~Hz, then the AD is cold.
The values of  $\nu_{0.01}\nu_{\rm max}$ depends on the AD SED, which is set by $a_*$. 
Table 1 provides $\nu_{0.01}/\nu_{\rm max}$ values for different $a_*$ values,
which all cluster at $\sim 3.5$. Thus, a non-ionizing AD peaks at $\nu_{\rm max}<9.4\times 10^{14}$~Hz,
or $\lambda_{\rm max}>3200$\AA. Since the typical x-ray luminosity observed is $L_x\la 0.1L_{\rm bol}$,
the ionizing luminosity of a cold AD is $\sim 10$ smaller than $L_x$, and the ionization will be
essentially all from $L_x$.

We can also define a weakly-ionizing AD as an AD where the ionizing luminosity is 
$\le 0.1L_{\rm bol}$. For such an accretion disk the AD ionizing luminosity is $\simeq L_x$.
Such objects will form a transition from UV dominated ionization to X-ray dominated ionization.
Table 1 lists the values of $\nu_{0.1}/\nu_{\rm max}$,
which cluster at $\sim 2$. Setting $\nu_{0.1}<3.29\times 10^{15}$~Hz for a weakly-ionizing AD
implies that in such an AD $\nu_{\rm max}<1.65\times 10^{15}$~Hz,
or $\lambda_{\rm max}>1800$\AA.

Using the calculated values of 
$\nu_{0.01}/\nu_{\rm max}$, $h\nu_{\rm max}/k\teff$, and $\teff/T_0$, 
for a given $a_*$, as provided in Table 1, we 
derive the following upper limits for a cold AD: 
\[ 
T_0\le 2.08\times 10^5~{\rm K} 
\]
for $a_*=0$, and
\[
 T_0\le 7.98\times 10^4~{\rm K}
\]
for $a_*=0.998$. Inserting these limits into eq. 5, yields the conditions
\begin{equation}
\mdot<3.42\times 10^{-3}m_8^2
\end{equation}
when $a_*=0$, and
\begin{equation}
\mdot<7.41\times 10^{-5}m_8^2.
\end{equation}
when $a_*=0.998$. If $m_8$ is known, then $\mdot$ can be estimated from the AD model fit to
the optical (5100\AA) luminosity, $L_{\rm opt}$. Using eq. 8 in Davis \& Laor 
(2011)\footnote{Note that there is an error in the numerical factors in eqs.~5 and 7 of 
Davis \& Laor (2011).  The numerical factor in eq. 5 should be $40/\pi^2(6 h G^2/5 c^2)^{1/3}$
rather than $160/\pi^3(6 \pi^2 h G^2 c^2/5)^{1/3}$.  The numerical factor in eq. 7 should be 
$1.4 \rm \; M_\odot \; yr^{-1}$ rather than $2.6 \rm \; M_\odot \; yr^{-1}$.  However, the
fitting function for $\Mdot$ in their eq. 8, which corresponds to our eq. 8, is unaffected.}
\begin{equation}
\mdot=3.5m_8^{-0.89}L_{\rm opt,45}^{1.5} ,
\end{equation}
where $L_{\rm opt,45}=L_{\rm opt}/10^{45}$, we obtain
the lower limits on $L_{\rm opt}$, below which the AD is cold,
\begin{equation}
L_{\rm opt,45}<9.85\times 10^{-3}m_8^{1.93}
\end{equation}
for $a_*=0$, and
\begin{equation}
L_{\rm opt,45}<7.66\times 10^{-4}m_8^{1.93}
\end{equation}
for $a_*=0.998$. Thus, even luminous quasars can become lineless, if $m_8$ is high enough.
For example, for $L_{\rm opt}=10^{46}$~erg~s$^{-1}$, the AD becomes too cold 
to ionize when 
$M>3.6\times 10^9\Msun$ for $a_*=0$, or for $M>1.4\times 10^{10}\Msun$ 
for $a_*=0.998$. The corresponding value for $l$ are $\la  0.22$ and $\la  0.06$,
using the approximation $L_{\rm bol}\simeq 10L_{\rm opt}$, i.e. 
$L_{\rm bol}=10^{47}$~erg~s$^{-1}$, and the 
implied relation $l=L_{\rm opt,45}/1.25m_8$.
Low luminosity AGN need to have lower $l$ values to have a cold AD.
For example, AGN at $L_{\rm opt}=10^{42}$~erg~s$^{-1}$ need to have 
$l\la  2.6\times 10^{-3}$ when $a_*=0$, and $l\la  7\times 10^{-4}$
when $a_*=0.998$, to become cold.

The value of $M$ is commonly estimated using the broad emission lines. 
Kaspi et al. (2000) gives
\begin{equation}
m_8=1.5L_{\rm opt,45}^{0.69}  v_{3000}^2,
\end{equation}
where $v_{3000}$=H$\beta$~FWHM/3000~km~s$^{-1}$.
Inserting the Kaspi et al. relation to eq. 8 above, gives 
(Davis \& Laor 2011, eq. 9)
\begin{equation}
\mdot =2.5 L_{\rm opt,45}^{0.87}  v_{3000}^{-1.78},
\end{equation}
Inserting these two expressions to eqs.~6-7 yields
\begin{equation}
v_{3000}\ge 2.72L_{\rm opt,45}^{-0.088} ,
\end{equation}
when $a_*=0$, and
\begin{equation}
v_{3000}\ge 5.28L_{\rm opt,45}^{-0.088} 
\end{equation}
when $a_*=0.998$. Thus, the criterion for a non-ionizing AD is set almost purely
by the value of the H$\beta$ FWHM, and is very weakly dependent on $L_{\rm opt}$. 
The AD becomes cold for H$\beta$ FWHM$>8160$~km~s$^{-1}$, 
when $a_*=0$, or for H$\beta$ FWHM$>15,840$~km~s$^{-1}$, when $a_*=0.998$. The 
quasar may become lineless when the H$\beta$ FWHM is above these limits. 
These maximal values for the H$\beta$ FWHM for an ionizing AD
may correspond to the maximum values of the observable H$\beta$ FWHM, 
as quasars with broader lines may drop out from quasar surveys which are based
on emission lines selection, due to their significantly weaker line emission.

Figure 1 shows the range of SEDs for various AD models. All models have
the same absolute accretion rate $\mdot=1$, and $\mu=0.8$. The upper two panels show 
local BB accretion disk models. In both panels the SED gets softer
as $M$ increases, since $\nu_{\rm max}\propto M^{-1/2}$ (see eq.~4). 
At a given $M$, the SED is softer for the $a_*=0$ model, and 
$\nu_{\rm max}$ is a factor of 2.4 lower compared to the
$a_*=0.998$ model (as can be deduced from Table 1). 
Note that although the $a_*=0.998$ AD is hotter than the $a_*=0$ case
by a factor of $\teff(a_*=0.998)/\teff(a_*=0)=4.5$,
the SED of the $a_*=0$ model has $h\nu_{\rm max}/k\teff=2.18$ 
compared to only 1.18 for the $a_*=0.998$ model, which makes the $a_*=0.998$ model
$\nu_{\rm max}$ only a 
factor of 2.4 higher than for the $a_*=0$ model.  Each curve is labeled with 
the ionization fraction $f_{\rm ion}$, which is defined as the fraction
of the bolometric luminosity that is radiated at all wavelengths shortward
of the the Lyman edge.

There are two effects which can make the observed spectrum harder than
shown in Fig.1 for the local BB models. These effects will
increase the required minimal H$\beta$ FWHM for an AD to become
cold. First, the AD SED is non isotropic. In the Newtonian case only
the normalization of the AD luminosity changes with inclination. However, the 
relativistic effects
change also the shape of the SED, making it harder closer to an edge on view.
The BLR must subtend an appreciable fraction of the sky, as seen by the ionizing
source. The BLR is most likely in the form of a thick torus-like configuration, coplanar
with the AD. The ionizing SED seen by the BLR will then be harder
than the observed SED, which is likely seen at a smaller inclination, closer to
a face on view. Assuming the torus-like configuration where the BLR resides, 
is restricted to $\mu<0.5$, and the clear line of sight cone where the AGN is unobscured
is restricted to $\mu>0.5$,
it is plausible to assume that the mean inclination of the BLR is 
$\mu=0.3$, while our line of sight has a mean value of $\mu=0.8$.
Table 1 provides $\nu_{\rm max}(\mu=0.8)/\nu_{\rm max}(\mu=0.3)$, and shows
that the BLR can be exposed to an SED a factor of $1.2-2.2$ harder than the observed SED. 
If we denote $x_{0.01}=h\nu_{0.01}/k\teff$, and
use the parameter $A_{\mu}=x_{0.01}(\mu=0.3)/x_{0.01}(\mu=0.8)$, then the minimal 
value for $v_{3000}$ 
derived in eqs.13,14 scales as $A_\mu^{0.69}$. Table 1 lists $A_{\mu}$  values 
for a range of $a_*$ values. 
Applying this inclination correction for the hardness of the ionizing SED 
implies minimal H$\beta$ FWHM values of  9090~km~s$^{-1}$
($a_*=0$), and 25,820~km~s$^{-1}$ ($a_*=0.998$).

The second effect which can make the AD SED harder is deviations from
the local BB approximation. A deviation from a local BB
necessarily leads to a photospheric temperature $T>T_{\rm eff}$. Table 2
provides the ratios of $\nu_{\rm max}/\teff$ derived from detailed AD
atmosphere models computed with TLUSTY.  These model are physically
equivalent to those presented in Hubeny et al. (2000), but utilize the
interpolation scheme described in Davis \& Hubeny (2006).  Since we
have a limited range of converged annuli, we only consider TLUSTY
models with $a_* \le 0.9$.  In the local BB AD models $\nu_{\rm
  max}/\teff$ depends only on $a_*$ and $\mu$, as $M$ and $\Mdot$ only
shift the SED, and do not affect its shape. In the atmospheric model
the local emission is set by the photospheric density and temperature, and
Table 2 provides models for the AD parameters which produce a relatively
cold AD, relevant for our study. Since $T>T_{\rm eff}$ the SED is
expected to be hotter compared to the BB model for the same
parameters, and thus produce a higher $x_{0.01}$ value. A notable
caveat is that some models have a strong Lyman absorption edge, which
can lead to a lower $f_{\rm ion}$ compared to the
local BB AD model, despite the higher $T$.

Figure 1, lower panel, presents the AD SED derived from the TLUSTY
models.  The deviation of these detailed calculations from the simple
BB approximation are strongly dependent on the local disk $\teff$ and
the photospheric density. The photospheric density also depends
(albeit more weakly) on the the stress prescription, which is
parametrized here using $\alpha=0.01$ (SS73). The SS73 model
gives a vertically averaged density which scales as $\alpha^{-1}$ in
the radiation dominated regime.  The dependence of the photospheric
density on $\alpha$ in TLUSTY models is typically much weaker, for
reasons discussed in the Appendix of Davis et al. (2006), unless
$\alpha$ and $\Mdot$ are both large enough that the disk begins to become
effectively optically thin. The SED can be significantly harder than
the local BB model when $\teff$ is high. For example, $\nu_{\rm max}$
increases by a factor $\sim 5$ for the $m_8=1$, $\mdot=1$
model. However, the significantly colder AD which are of interest
here, are not far from the local BB approximation.  Table 2 provides
the correction factors $A_{\rm model}=x_{0.01}(\rm
TLUSTY)/x_{0.01}(\rm BB)$ for $\mu=0.8$. Since $v_{3000}\propto
A^{0.69}$, atmospheric effects can increase the minimal H$\beta$ FWHM
values by a factor of $\sim 1.2-2$.

\section{Observations}

The SED of quasars with a cold AD should show 
a sharp cutoff beyond the Lyman limit. The WLQ population comes closest 
to the lineless quasars population. The WLQ with the highest rest frame UV energy
probed is SDSS J094533.99+100950.1, at $z=1.66$, discovered by Hryniewicz et al. 
(2010). The GALEX FUV photometry extends down to rest frame 570\AA.
Figure 2 presents a local BB AD match to the overall SED presented by 
Hryniewicz et al. The AD model parameters were: $m_8=27$, as estimated by 
Hryniewicz et al. based on the Mg~II FWHM and the 3000\AA\ continuum luminosity.
We assume an AD with a face on view ($\mu=1$). The value of $\Mdot$ was 
then adapted to reproduce the overall SED normalization. The remaining 
free parameter is $a_*$, which is varied to match the FUV turnover. A 
suitable match to the data is found with $\mdot=11.4$ and $a_*=0.3$. A good match can 
also be obtained with a higher inclination AD model, where the resulting 
harder SED is compensated by allowing a lower $a_*$. A reasonable match 
can be obtained down to $\mu=0.6$, and $a_*=-1$, i.e. a BH counter-rotating 
with respect to the AD.
A higher inclination, $\mu<0.6$, is excluded, as the
AD SED becomes too hard, and $a_*$ cannot be further reduced to compensate for that.
Similarly, a value of $a_*>0.3$ is excluded, as the AD is too hard and 
and $\mu$ cannot be further increased to compensate for that. These results are similar
to those obtained by Czerny et al. (2011), who also discuss the effects of
intrinsic reddening on the derived AD parameters. 
We also attempted TLUSTY model fits to the data, however the SED of these 
models is broader than the local BB AD models, and a fit of comparable quality 
to the local BB AD models fit could not be obtained.

Could the observed UV emission be significantly affected by the foreground
Lyman forest absorption? Barger \& Cowie (2010, Fig.6 there) show that the
cumulative effect of the Lyman $\alpha$ absorption systems along the line
of sight, suppresses the continuum at rest frame 912\AA$<\lambda<$1216\AA\
by a level of 4\% for $z=1.1$.  Using the evolution of the 
Lyman forest systems, $dn/dz\propto (1+z)^{2.23\pm 1.2}$ for $0.9<z<1.7$
(Janknecht et al. 2002),
the expected continuum suppression at $z=1.66$ is $<11$\%, and is thus 
not significant. However, Lyman limit systems can produce 
significant absorption at rest frame $\lambda<$912\AA. To affect the observed 
SED the Lyman limit absorber should occur longward of observed 1516\AA\ 
(the GALEX FUV effective wavelength), i.e have $z>0.66$. Using the 
number density per unit redshift evolution of $dn/dz=0.15\times (1+z)^{1.9}$
(Songaila \& Cowie 2010), the expected number of absorbers 
integrated over $0.66<z<1.66$, is 0.66. A Lyman 
limit absorber will likely cause a break in the spectral curvature, which is not seen
in the data (Fig.2). A further complication may arise if a foreground
Lyman break is diluted by a Lyman continuum emission edge produced by the AD.
UV spectroscopy is required to clearly exclude 
foreground absorption
effects on the observed FUV SED of SDSS J094533.99+100950.1.

\section{Discussion}

The clear prediction of local BB AD models is that AGN with the correct
combination of $M$ and $l$ harbor
a luminous, but non-ionizing, continuum source. The UV SED of an AGN with
$f_{\rm ion}=0.01$ will show
a peak at $\lambda>3200$\AA, and a steep drop at $\lambda<1000$\AA, and may
form a lineless quasar, depending on the
relative strength of the X-ray ionizing power-law continuum.
An AGN with $f_{\rm ion}=0.1$ will show a peak at 
$\lambda>1800$\AA, and will likely form a WLQ, with an SED similar to that observed in
the WLQ SDSS J094533.99+100950.1.

A cold non-ionizing AD is expected in luminous,
$L_{\rm bol}\ga  10^{47}$~erg~s$^{-1}$, quasars even at
relatively high values of $l\la  0.2$, in particular for $a_*\simeq 0$.
In low luminosity AGN, $L_{\rm bol}\la  10^{43}$~erg~s$^{-1}$, the SED
is predicted to become non-ionizing only at very low values of $l\la  10^{-3}$.
Since the X-ray/optical luminosity ratio is observed to increase with decreasing $L_{\rm bol}$,
such low luminosity and low $l$ AGN may be characterized by more prominent 
line emission, excited by the relatively stronger X-ray power-law emission.

What is the nature of the expected BLR emission? The flat X-ray power-law
continuum produces an extended region of low ionization in the BLR gas, 
which is expected to cool
mostly through collisionally excited low ionization lines, such as Fe~II, Mg~II,
and Balmer lines (collisionally excited due to highly trapped Ly$\alpha$ photons).
The narrow line region gas is expected to
cool mostly through forbidden low ionization lines, such as
O~I, N~II, and S~II lines, as seen in LINERs (see photoionization
model results in Ferland \& Netzer 1983).

\subsection{Are we missing lineless quasars?}
Is it possible that AGN surveys are missing a
significant number of moderately high $l$, high $M$ AGN which are lineless?.
This can be addressed by
surveys of AGN based on color, independent of line emission. This was carried out by
Diamond-Stanic (2009) using the SDSS sample, where they defined WLQ as having
a Ly$\alpha$ EW$<$15.4\AA, which is 3$\sigma$ from the mean Ly$\alpha$ EW=63.6\AA.
They find that the fraction of quasars classified as WLQ increases from
1.3\% at $z<4.2$ to 6.2\% at $z>4.2$. Thus, lineless quasars do not form a
significant part of the quasar population, as expected given the required 
very high $M$ values in luminous AGN, in order to have a cold AD.

\subsection{Do LINERs have a cold AD?}
In contrast with luminous AGN, a significant fraction of 
low luminosity AGN near the more massive black hole should 
fall in the cold AD 
regime. Theoretically, it is not clear whether the thin AD configuration 
remains valid at $l\la  10^{-2}$, 
or whether the accretion transforms 
to an optically thin and geometrically thick configuration, which may produce
a hard power-law continuum source (e.g. Narayan et al. 1998). Observationally, 
the narrow emission lines of low luminosity AGN in massive galaxies
are generally characterized as LINERs, and are likely to be powered by a hard
continuum source (Ferland \& Netzer 1983), as their observed SED apparently suggests
(e.g. Ho 2008, Eracleous et al. 2010a; cf. Maoz 2007). 
Thus, the significant contribution 
of X-ray heating in low luminosity AGN, given their flat $\alpha_{ox}$, may imply
that FUV heating is not significant in these objects anyhow, and its absence as the AD becomes 
cold will
not affect their line emission significantly, unlike typical luminous AGN where the FUV
dominates the line excitation. The line emission in LINERs may also be powered by 
additional mechanisms (e.g. Eracleous et al. 2010b), which further lowers the diagnostic
power of the line emission on the ionizing SED shape.
 
Interestingly, Lawrence (2005) finds that the change in the SED of a few 
potentially characteristic AGN, extending over a range of $10^6$ in luminosity,
is consistent with the expected change for an optically thick AD at a fixed $M$. 
It remains to be explored whether the optical-UV SED in the lowest luminosity
AGN (e.g. LINER) can be fit by a cold AD,
or whether only a featureless hard power-law is required. A cold AD in low luminosity
AGN cannot produce a lineless AGN, unless $\alpha_{ox}$ is unusually steep, and
ionizing radiation is indeed missing. 

\subsection{True type 2 AGN}
In ``true type 2 AGN'' which show normal (non LINER) narrow lines, the absence of 
broad lines does not result from a cold AD, as an apparently normal photoionizing 
radiation, which excites the narrow lines, is present.  In such objects, the BLR
may be absent due to the absence of the mechanism which produces it, such as an AD wind, 
as proposed by Nicastro (2000), 
Nicastro et al. (2003), Czerny et al.(2004), Elitzur \& Shlosman (2006) and
Elitzur \& Ho (2009).

\subsection{What produces WLQ?}
Are most WLQ powered by a cold AD?  The archetype WLQ, PG~1407+265, is probably not,
as it's optical-UV SED appears to be similar to the average quasar SED (McDowell et al. 1995). 
It does not show the expected peak at $\lambda>1800$\AA\ and a sharp drop at 
$\lambda<1000$\AA\ (although the the S/N at $\lambda<1000$\AA\ is rather low).
However, its peculiar emission 
line properties, with some detectable low ionization lines (Balmer lines, Mg~II, Fe~II), 
but very weak or no detectable higher ionization lines, is characteristic of other WLQ 
(Leighly et al. 2007a; Diamond-Stanic et al. 2009; Plotkin 2010b; Shemmer et al. 2010; 
Hryniewicz et al. 2010; Wu et al. 2011). In the case of PHL~1811, this line emission
is modeled by an unusually soft ionizing spectrum (Leighly et al. 2007b), 
which is however qualitatively similar to the emission of a cold AD. The soft ionizing 
SED produces little heating per ionization, and thus weak collisionally 
excited lines compared to recombination lines.

Some WLQ may be driven by a different processes. The C~IV EW is tightly inversely related 
with $l$ (Baskin \& Laor 2004), and also with other emission line properties, such as the
[O ~III] and Fe~II EW (Baskin \& Laor 2005), which are part of the eigenvector 1 correlations
of various emission line properties (Boroson \& Green 1992). The C~IV EW is also
related to the C~IV line peak velocity shift and profile asymmetry 
(e.g. Baskin \& Laor 2005; Richards et al. 2011), and $\alpha_{ox}$ 
(Baskin \& Laor 2005; Wu et al. 2009).
This large set of correlations suggests that local effects 
in the BLR (e.g. metallicity, kinematics), 
may control the EW of C~IV, and possibly of other high ionization UV lines. 
If the driving mechanism for WLQ 
is a filtered ionizing continuum which illuminates the BLR (e.g. Richards et al. 2011; 
Wu et al. 2011), then
the effect on the BLR excitation may be similar to the effect of a cold AD. However, the trend
of decreasing C~IV EW with increasing $l$ cannot be confused with the cold AD effect,
as higher $l$ implies a hotter AD SED, rather than a colder one.

\subsection{The absence of AGN with very broad lines}
The combination of $M$ and $\Mdot$ which produces a cold AD, can be transformed to
a limit on the H$\beta$ FWHM, with very weak $L_{\rm opt}$ dependence. 
The local BB AD may become cold at
 $v> 8,000$~km~s$^{-1}$ when $a_*=0$, and for $v> 16,000$~km~s$^{-1}$ 
when $a_*=0.998$. The lower limit on $v$ can be higher by a factor of up to 2 
due to the inclination
dependence of the ionizing continuum, and possible deviation of the disk emission 
from a local BB continuum.
AGN do show a sharp drop in their number at 
$v\ga  10,000$~km~s$^{-1}$ (e.g. Hao et al. 2005, Shen et al. 2008), 
and this drop is independent of luminosity (Shen et al. 2008, Fig.3 there).
Laor (2003) suggested there
may be a yet unknown physical mechanism which suppresses the BLR at $v>25,000$~km~s$^{-1}$.
The AD becoming too cold to ionize the BLR, may provide a natural explanation for the observed 
luminosity independent limiting $v$, beyond which broad lines are not seen.

How does the line emission of AGN with the broadest lines appear?  Do such objects 
enter the WLQ regime? Such a systematic study is not available yet. However,
very broad line objects tend to have a double-peaked emission profile, and the
spectrum of the ``prototypical'' such object, ARP~102B, shows strong Balmer and Mg~II 
lines, strong low ionization forbidden lines, and relatively weak higher ionization 
lines (Halpern et al. 1996). Only anecdotal information is available for a couple
of more such objects (Storchi-Bergmann et al. 2005; Eracleous et al. 2009), partly 
consistent with the results for ARP~102B. A systematic study is clearly required
to explore whether very broad line AGN indeed have emission line properties
intermediate between normal AGN and WLQ. 

What is the SED of very broad line quasars?
Strateva et al (2008) shows the overall SED of five double-peaked Balmer lines
AGN, with $v\sim 14,000-20,000$~km~s$^{-1}$. Their optical-UV SED is generally
consistent with the average AGN SED, but the FUV at $\lambda<1000$\AA\ is not
probed. The objects have a flatter $\alpha_{ox}$ than the average for quasars at a similar 
luminosity (see also Strateva et al. 2006). This may indicate indirectly that
significant X-ray heating is required for generating the broad lines in these objects,
as very broad line quasars with an average $\alpha_{ox}$ may be selected against due to their weaker
line emission. The flatter $\alpha_{ox}$ may therefore compensate for a smaller 
contribution of the UV continuum, if their AD is indeed colder. A systematic study
of the FUV emission of very broad line AGN can test if their FUV is indeed steeper.

\subsection{Continuum based searches for lineless quasars}
An alternative approach to test the cold AD scenario is to explore the emission line 
properties of AGN where the SED indicates weak FUV emission. It is important to note
that red quasars (e.g. Richards et al. 2003) are not suitable objects, as a cold AD
produces a normal blue continuum at $\lambda>3200$\AA. Red quasars are either affected
by dust extinction, or by some other modification of the continuum emission mechanism.
Cold AD candidates can be found by looking for quasars with a blue NUV slope,
$\alpha_{\lambda>3200{\rm \AA}}>-1$, and a red FUV slope, $\alpha_{\lambda<1000{\rm \AA}}<-2$,
as seen in SDSS J094533.99+100950.1 (Fig.2). Such observation cannot be achieved from
the ground, as the required $z>3$ to probe the FUV slope from the ground,
leads to a high probability for an intervening Lyman limit absorption
system ($dn/dz>2$). Space based UV observation of $z<2$ quasars are essential for that, 
as was done by Hryniewicz et al. (2010) using GALEX.

The observed optical-UV SED of optically selected quasars shows a small dispersion 
(Elvis et al. 1994; Richards et al. 2006). Due to this
Davis \& Laor (2011) deduced that the accretion efficiency, $\eta$, increases with $M$,
to compensate for the drop in the FUV with increasing $M$ for AD models at a fixed $\eta$.
One may worry that the implied rise in $\eta$ with $M$ is just a selection
effect, as high $M$ and low $\eta$ quasars necessarily have a cold AD, produce weak line emission,
and are selected against in a broad emission line AGN survey. However, a color based
AGN survey is not blind to a cold AD quasar, and as noted above,
Diamond-Stanic (2009) find that only 1.3\% of $z<4.2$ color selected AGN are WLQ. 
Thus, there is no large population of low $a_*$ high $M$ quasars, and the high
mean $\eta$ (high $a_*$) values for high $M$ AGN deduced by Davis \& Laor (2011), 
is not driven by emission line selection.

We note that quasars generally show a soft-excess X-ray emission below 1~keV,
compared to the 2-10~keV power-law emission (Wilkes \& Elvis 1987), 
which may extend as a single power-law component to the FUV (Laor et al. 1997).
This component is commonly attributed to comptonization in a warm surface layer 
above the AD (e.g. Kriss et al. 1999). Such a layer
can turn a cold AD SED into an ionizing SED. The SED of 
SDSS J094533.99+100950.1 is consistent with the exponential cold AD FUV cutoff, and
shows no indication for a power-law component. Such a warm comptonizing surface
corona may be lacking in a cold AD (Czerny et al. 2011).

\subsection{Cold AD in Blazars}
Ghisellini et al. (2009, 2010) attribute the optical-UV emission
feature in blazars observed with the {\em Swift} satellite to thermal AD emission, 
based on the low polarization and low variability of this component. This component 
shows a peak in the NUV and a steep falloff in the FUV, and is fit with an AD model 
with very high values of $M$, up to $4\times 10^{10}\Msun$ in some objects. 
These are therefore generally cold AD models, and if the BLR emission in these objects is 
powered mostly by the unbeamed thermal disk emission, some of these blazars should appear
as WLQ. Indeed, some of the objects in Ghisellini et al. (2010) do show weaker
UV lines (e.g. S5 0014+81, see Kuhr et al. 1983, Sargent et al. 1989; 
PKS~2149-306, see Wilkes 1986; 
RBS~315, see Ellison et al. 2008), which may provide independent support for the cold AD 
interpretation of their optical-UV bump.
However, intervening Lyman limit absorption systems are clearly contributing to 
the observed FUV steep falloff in these $z\sim 2.5-3.5$ blazars, so the intrinsic 
AD continuum may not be as cold as observed. In addition, one cannot
exclude additional line excitation from the highly luminous jets in these objects.

\section{Conclusions}

Standard thin AD models imply a non ionizing continuum for high $M$ 
($\ga  3\times 10^9\Msun$), even for
fairly luminous quasars ($L_{\rm opt}=10^{46}$~erg~s$^{-1}$), 
in particular if the black hole is non-rotating. Such cold AD may reside in
AGN where the Balmer line FWHM$> 8,000-24,000$~km~s$^{-1}$.
The ionizing continuum in such objects originates only in the X-ray
power-law component, which will likely produce an extended partially ionized
region in the illuminated gas, which cools mostly through low ionization lines.
If $L_x\la  0.1L_{\rm bol}$, as commonly
seen in luminous AGN, the ionizing luminosity will be weaker by a factor of $\ga  5$
than in average AGN. Such objects will appear as WLQ. If $L_x<< L_{\rm bol}$ such
quasars may be lineless.

Cold AD provide a natural explanation for the sharp drop in the number of AGN with 
broad line FWHM$\ga  10,000$~km~s$^{-1}$, which is independent of luminosity. 
Followup FUV observation are
required to test if very broad line AGN show the expected spectral turnover
in the FUV, if they are powered by a cold AD. The best strategy is to
obtain UV spectra of $z\sim 1-2$  very broad line quasars, to minimize the
likelihood of intervening Lyman limit absorption by the IGM, or to detect
such absorption if it occurs.

Color selected AGN samples indicate that only a small fraction ($<6$\%)
of luminous AGN may harbor a cold AD. This can result from the high
mean $\eta$ values for AGN at high $M$, which may imply a higher $a_*$ 
and a hotter AD emission which produces a higher $f_{\rm ion}$.

Searches for quasars with a blue optical-NUV and a red FUV continuum,
using ground based optical spectra, and space based UV spectra of
$1< z\la 2$ quasars, may be an efficient
mechanism to reveal cold AD AGN. The continuum of a WLQ provided by
Hryniewicz et al.  (2010), which extends down to $\lambda=570$\AA, is
surprisingly well fit by a simple local BB AD models. 
This is in
contrast with most AGN, where the SED cannot be well fit by a simple
local BB AD model. The local BB models provide a better match than the
TLUSTY-based models (Hubeny et al. 2000), suggesting that the
effects of the Lyman edge and electron scattering opacity on the
spectrum may be overestimated by these models.

Since both $M$ and $L_{\rm opt}$ are determined from the observations
of SDSS J094533.99+100950.1, only $a_*$ and $\mu$ remain
as free parameters to match the observed SED. The fit allows us to place an upper limit on the
black hole spin ($a_*\leq 0.3$), and a lower limit on the inclination of the
AD ($\mu>0.6$).  The precise limits are somewhat uncertain, in part
because the Mg~II line-based estimates of $M$ are uncertain.  A larger
$M$ would allow a wider range of $a_*$ and $\mu$, but a smaller $M$
would provide tighter constraints: an even lower limit on $a_*$ and
higher limit on $\mu$, as further discussed by Czerny et al. (2011).

It is worthwhile to explore whether there are additional objects which can
be well fit by a cold AD SED, with only $a_*$ and $\mu$ as free parameters.
If such objects exist, the SED may provide
a useful tool to constrain $a_*$ and $\mu$ in such AGN, as commonly done
in X-ray binaries. The AD fit may also provide some new insight on the 
structure of the accretion disk atmosphere.

\section*{Acknowledgments}

We thank K. Hryniewicz for providing the data in electronic form. 
We thank Ohad Shemmer for useful discussions, and the referee for useful comments.
SWD is supported in part through NSERC of Canada.

{}

\begin{figure*}
\begin{center}
\psfig{figure=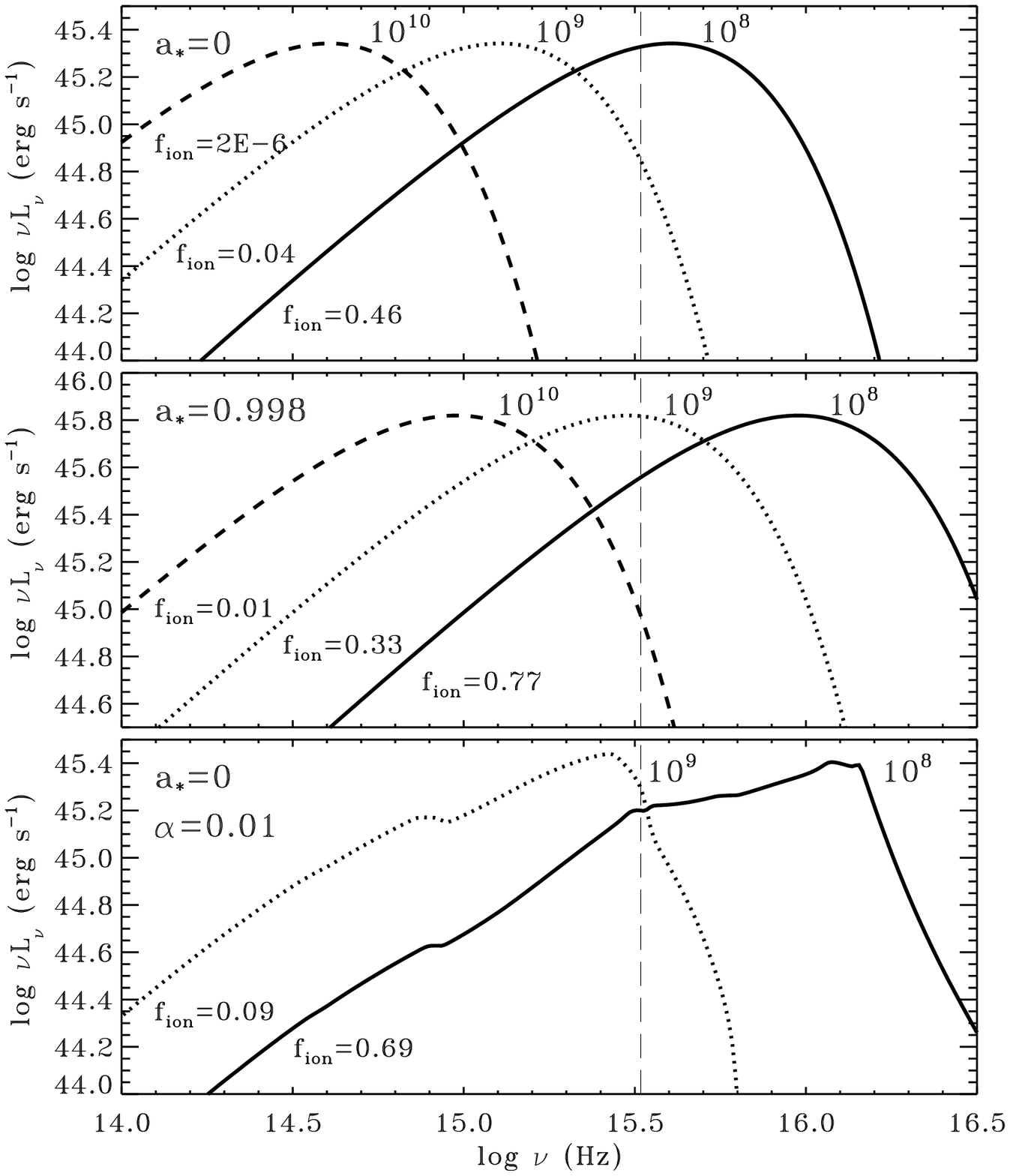,width=0.95\textwidth,angle=0}
\end{center}
\caption{The SED of various AD models. All models have 
$\mdot=1$ and $\mu=0.8$. The vertical dashed line shows
the threshold $\nu$ for ionizing H. The two upper panels show various 
local BB accretion disk models for $a_*=0$ and $a_*=0.998$. The SED gets softer
as $M$ increases, and as $a_*$ decreases.
The lower panel shows the TLUSTY model (Hubeny et al. 2000). 
with $\alpha=0.01$. The SED is harder than the local BB model, and the effect
is larger when $\teff$ is larger.
The fraction of ionizing radiation, $f_{\rm ion}$, is raised by the atmospheric effects,
compared to the local BB AD models.}
\end{figure*}

\begin{figure*}
\begin{center}
\psfig{figure=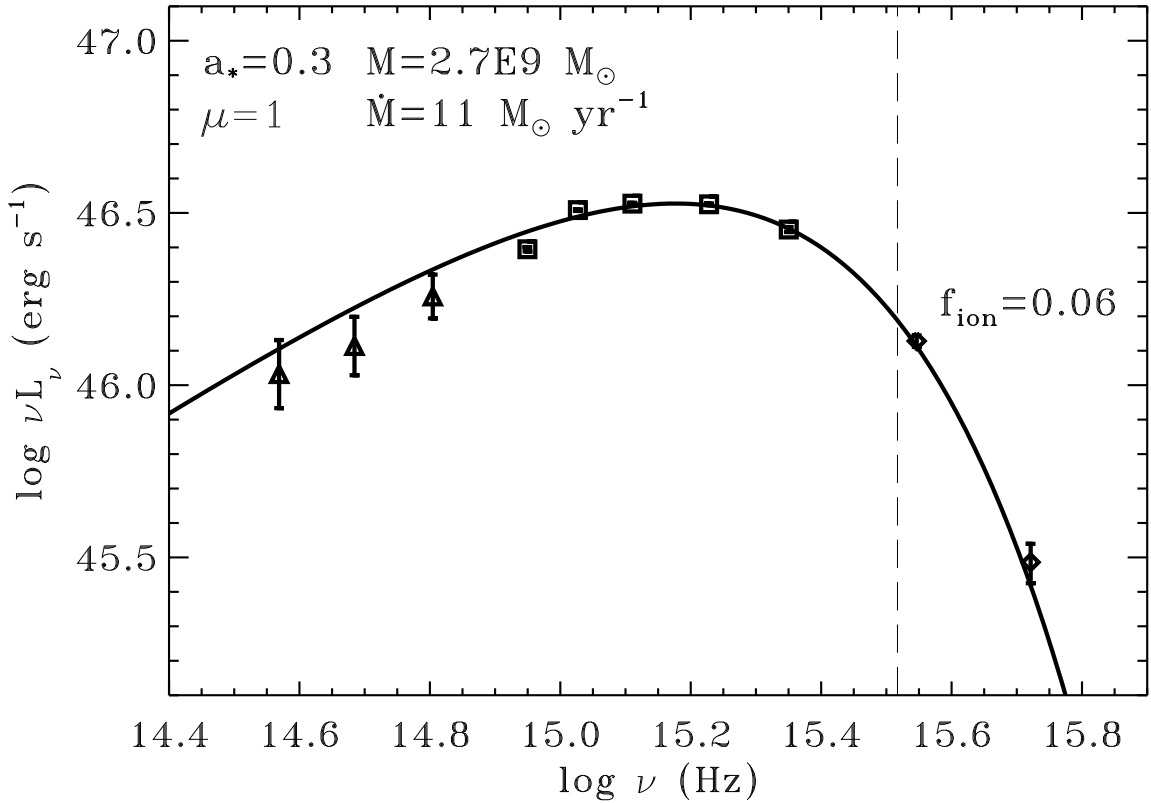,width=0.95\textwidth,angle=0}
\end{center}
\caption{ 
A local BB AD model match to the $z=1.66$ weak emission line quasar SDSS J094533.99+100950.1 
from Hryniewicz et al. (2010). The TLUSTY models produced a lower quality match to the data.
The photometric data points are taken from non simultaneous
observations by 2MASS (triangles), SDSS (squares), and GALEX (diamonds).  The AD model
fit parameters are listed. The value of $M$ is taken from Hryniewicz et al. 
The model peak position is mostly set by the combination of $\mu$ and $a_*$, and the
normalization mostly by $\Mdot$. The AD model fit gives $f_{\rm ion}=0.06$,  
above our definition of $f_{\rm ion}=0.01$ for potentially lineless quasar, and this higher value 
may be driving the weak low ionization line emission observed.}
\end{figure*}

\newpage
\begin{table*}
 \begin{minipage}{140mm}
  \caption{}
  \begin{tabular}{lcccccc}
  \hline
   $a_*$     &  $\teff/T_0$ & $h\nu_{\rm max}/k\teff $ & $\nu_{0.1}/\nu_{\rm max}$ &
$\nu_{0.01}/\nu_{\rm max}$ & $\nu_{\rm max}(\mu=0.8)/\nu_{\rm max}(\mu=0.3)$ &
$A_\mu$ \\
    \hline
0 & 0.1 & 2.18 & 1.91 & 3.43 & 0.85 & 1.17\\
0.9 & 0.23 & 1.77 & 1.97 & 3.54 & 0.64 & 1.59\\
0.998 & 0.45 & 1.18 & 2.03 & 3.54 & 0.46 & 2.03\\
\hline
\end{tabular}
\medskip

The results for local BB AD models, observed at $\mu=0.8$. All the results in the table are independent of the values of $m_8$ and $\mdot$.
\end{minipage}
\end{table*}

\begin{table*}
 \begin{minipage}{90mm}
  \caption{}
  \begin{tabular}{lccccc}
  \hline
$a_*$ & $m_8$ & $\mdot $ & $h\nu_{0.1}/k\teff$ & $h\nu_{0.01}/k\teff$ & $A_{\rm model}$ \\
    \hline

0 & 10 & 1.22 & 6.1 & 9.9 & 1.32\\
0 & 32 & 12.2 & 6.8 & 11.6 & 1.56\\
0 & 100 & 123 & 7.2 & 16.0 & 2.15\\
0.9 & 10 & 0.28 & 6.6 & 12.1 & 1.93\\
0.9 & 32 & 2.84 & 8.3 & 14.5 & 2.32\\
0.9 & 100 & 28.4 & 8.9 & 16.6 & 2.67\\
\hline
\end{tabular}
\medskip

The TLUSTY model results, observed at $\mu=0.8$.
All models have a fixed $m_8\times \mdot=1$, and therefore a fixed
$T_0$. For the BB model, $h\nu_{0.1}/k\teff=4.2-3.5$,
and $h\nu_{0.01}/k\teff=7.5-6.3$, for $a_*=0-0.9$. In contrast
to the BB case, we do not compute $a_*=0.998$ models because our 
TLUSTY model atmosphere table is not extensive enough to reliably 
compute disk spectra for such high spins.
\end{minipage}
\end{table*}

\label{lastpage}

\end{document}